# A chemical avenue to manipulate field-reentrant superconducting competition in infinite layer nickelates

Haowen Han[1,†], Yusong Zhao[1,†], Yi Bian[1,†], Tong Ma[1], Wenlong Yang[2], Shaohua Yang[3], Binghui Ge[3,*], Hongliang Dong[4,5], Chuanying Xi[6], Ze Wang[6], Nuofu Chen[7], Tian Shang[8], Toni Shiroka[9], Zaher Salman[9], Jia-Cai Nie[2,10,*], Ho-Kwang Mao[4,5], and Jikun Chen[1,*]

[1]Beijing Advanced Innovation Center for Materials Genome Engineering, School of Materials Science and Engineering, University of Science and Technology Beijing, Beijing 100083, China.

[2]School of Physics and Astronomy, Beijing Normal University, Beijing 100875, China.

[3]Institute of Physical Science and Information Technology, Anhui University, Anhui 230601, China.

[4]Center for High Pressure Science and Technology Advanced Research, Shanghai 201203, China.

[5]Shanghai Key Laboratory of Material Frontiers Research in Extreme Environments, Institute for Shanghai Advanced Research in Physical Sciences, Shanghai 201203, China.

[6]Anhui Key Laboratory of Low-Energy Quantum Materials and Devices, High Magnetic Field Laboratory, HFIPS, Chinese Academy of Sciences, Anhui 230031, China.

[7]School of Renewable Energy, North China Electric Power University, Beijing 102206, China.

[8]Key Laboratory of Polar Materials and Devices (MOE), School of Physics and Electronic Science, East China Normal University, Shanghai 200241, China.

[9]PSI Center for Neutron and Muon Sciences, Villigen PSI 5232, Switzerland.

[10]Key Laboratory of Multiscale Spin Physics, Ministry of Education, Beijing Normal University, Beijing 100875, China.

***Corresponding authors**. E-mails: jikunchen@ustb.edu.cn; jcnie@bnu.edu.cn; bhge@ahu.edu.cn

[†]Equally contributed to this work.

**Abstract**

Recently, preliminary magnetic field-reentrant superconductivity manifested in high-temperature ($T_c$) Eu-doped infinite-layer (IL) nickelates, beyond analogous discoveries exclusively in low-$T_c$ systems. This evokes intriguing fundamental issues about potential quantum-phase boundary and criticality between unconventional superconductivity and field-reentrant-one, which are inexplicable owing to formidable challenges in growing IL-nickelates towards later-series rare-earths. Herein, we open up chemical avenues to enable effective growth of $(RE_{1-y}RE'_y)_{1-x}Eu_xNiO_2$ (*RE*/*RE'*: Pr, Nd, Sm, Gd, Dy), giving rise to discoveries of *RE*-4*f*-related quantum competition between high-$T_c$ and reentrant superconductivity. Robust magnetic-field-reentrant superconductivity with uniaxial anisotropy is observed at superconducting-dome boundaries, stemming from $Eu^{2+}$-$4f^7$ associated competition between magnetic-fluctuation promoted pairing and exchange-field interactions. Their quantum-criticality is further modulable via *RE*(*RE*')-magnetism, which either reinforces reentrancy or elevates $T_c$ (~40.1 K) with more robust critical-current-density (~266 kA/cm$^2$ at 2 K) beyond Sr-/Ca-doped counterparts. Our synthetic route enables the establishment of an ideal platform via IL-nickelates for studying 4*f*-related unconventional superconductivity and quantum-criticality.

## INTRODUCTION

The participance of *f*-electrons in ferromagnetic coupling[1], Kondo lattice[1,2], and antiferromagnetic exchange mediated Cooper-pairing[3,4], largely enriches the superconducting phase diagram, also giving rise to magnetic field-reentrant superconductivity[5]. Recently, preliminary sign of reentrant superconductivity was observed via introducing the half-filled $Eu^{2+}$ ($4f^7$) as hole dopant for infinite-layer (IL) nickelates[6-8], which belongs to a new family of high temperature superconductor[9] with $T_c$ near ~40 K[10]. This is in stark contrast to the analogous earlier discoveries in exclusive low-$T_c$ systems, such as $UTe_2$[5,11,12], URhGe[13], Eu-containing Chevrel phase compounds[14,15], λ-$(BETS)_2FeCl_4$[16,17] and moiré graphene[18]. It unveils a previously unexplored dimension associated with the rare-earth (*RE*) 4*f*-orbital and/or magnetic effects that may exert a substantial effect on superconducting phase diagram of nickelates, beyond their conventional ionic size effects[19,20]. This observation further coincides with the distinct magnitude and anisotropy in the superconducting upper critical field as observed for IL nickelates with various magnetic contributions by the *RE*-4*f* moments[21]. Also, it is more intriguing to note the generally elevated $T_c$ of ambient pressured nickelate superconductor via substituting their *RE* composition towards later lanthanide series, as presently valid for not only IL-nickelate[10,21,22], but also thin film layered-perovskite nickelates[6,23-26]. In light of the prevailing trajectory of nickelate superconductors, venturing into heavier *RE* to map potential 4*f*-orbital effects on superconducting phase diagram offers the prospect of groundbreaking superconductivities and fundamental elucidations of underlying mechanisms.

Nevertheless, the key challenge in exploring nickelate superconductors towards heavier-*RE* lies in their material growths[10,22,27]. Presently, successful growths of IL-nickelates manifesting superconductivity were exclusively via vacuum epitaxy of perovskite nickelates precursors followed by topotactic reduction, limited to light-*RE* prior to Eu[27]. Owing to the lanthanide contraction, introducing heavier *RE* (e.g., behind Eu) with smaller ionic radius of *RE* ($r_{RE}$) into perovskite nickelates is thermodynamically more difficult. This is because that a more distorted $NiO_6$ octahedra elevates the formation free energy ($\Delta G$)[19] of perovskite nickelates towards more positive magnitude, which is hardly stabilizable by coherent lattice with the substrate. Further obstacle is from the restriction in compositing the alkaline-earth (*AE*) hole dopants within perovskite nickelates, since conventional chemical process is incapable to form $Ni^{4+}$ in perovskites as major constituents even at extremely high $p_{O2}$ of ~GPa[20,28]. Thus, perovskite nickelates containing *AE* constituents were likely to be heterogeneously formed via plasma or atomic beam interplays with the substrate, contingent upon precise and narrowly defined experimental window[10,22,27]. From these perspectives, the present strategy for growing IL-nickelates is insufficient to support further explorations pertaining to heavier *RE*, despite its strong likelihood to realize higher $T_c$ and/or unconventional superconductivity, e.g., field-reentry[6-8].

Here, we open up a simple high-$p_{O2}$ assisted chemical avenue for growing IL nickelate superconductors toward heavier *RE* constituents with effectiveness in *RE*-substitutions, grounded in which their 4*f*-orbital relevant field-reentrant superconducting phase diagram is elucidated. Robust uniaxially anisotropic superconducting reentrancy was validated to emerge at the quantum phase boundaries of the superconducting dome for both $Nd_{1-x}Eu_xNiO_2$ and $Pr_{1-x}Eu_xNiO_2$ systems. Their quantum criticality was further modulated via introducing *RE'* within $(Nd_{1-y}RE'_y)_{1-x}Eu_xNiO_2$ (*RE'*: Pr, Sm, Gd, Dy and $Sm_{1/2}Pr_{1/2}$) to adjust the magnitude of exchange field associated with *RE*-magnetism. We highlight the further elevation in $T_c$ and critical current density ($J_c$) for the Eu-doped IL-nickelates, exceeding previous layer spacing expectations developed from their counterpart systems doped by Sr/Ca. The root-cause was further elucidated, combining characterization of pairing strength via magneto-transport and localized magnetism via low-energy muon-spin spectroscopy (LE-$\mu$SR). Our work unveils the pivotal role of magnetic fluctuations with close associations with *RE*-4*f* in modulating the quantum criticality and pairing strength, beyond conventional BCS theory. Also, our MPa-$p_{O2}$ assisted chemical synthetic route sheds a light on the capability to apply IL-nickelates as coated conductors analogous to high-$T_c$ cuprates[29].

## RESULTS AND DISCUSSIONS

### A MPa-high $p_{O2}$ assisted chemical avenue enables effective growth of IL-nickelates

To reduce the positive $\Delta G$ of perovskite nickelates contracting with lanthanide contractions, we exploit a MPa-high $p_{O2}$ assisted chemical strategy for thin film growth of IL-nickelate, illustrated in Fig. S1a. In brief, the chemical precursors of $RE(NO_3)_3$ and $Ni(CH_3COO)_2$ were dissolved in ethylene glycol monomethyl ether (EGME), and spin coated on a $NdGaO_3$ (110) substrate. The spin coated films were crystallized into the perovskite precursor film by annealing at 700-900 °C under a high-$p_{O2}$ within 1-20 MPa. Afterwards, the perovskite nickelates was transformed into infinite layer via soft chemical topotactic reduction based on $CaH_2$ co-anneals at a temperature between 260-320 °C for 0.5-4 hours. Instead of using conventional alkaline earth elements (e.g., Sr or Ca)[30,31], herein the hole doping was realized via partially substituting their *RE*-constituents by Eu, which displays variable valance state from +3 in perovskites towards +2 upon topotactic reduction[22]. From the thermodynamic perspective, the metastable perovskite nickelates precursors containing later series *RE*-composition is preferentially stabilized at a high-$p_{O2}$ within $10^0$-$10^2$ MPa[19,20], as indicated by their equilibrium phase chart in Fig. 1a (see more discussions in Figs. S1-S3). Hence, our strategy is capable to introduce later series *RE* further behind Eu into IL-nickelates, while the *RE*-constituent is simply and flexibly modulable manipulating the chemical solutions for spin coating.

Efficient growth of IL-nickelates is firstly exemplified by $Nd_{1-x}Eu_xNiO_2$, covering a large variety in Eu-compositions. Via MPa-high $p_{O2}$ anneals, the metastable $Nd_{1-x}Eu_xNiO_3$ is well stabilized, as indicated by their abrupt metal-insulator transitions (see Fig. 1b and Fig. S4)[32,33]. The accuracy in the stoichiometry controls for Eu is convinced by the linear increase in metal-insulator transition temperature ($T_{MIT}$) with $x$ (or average $r_{RE}$) in consistency with the previous reports[19,20], as shown by the inset of Fig. 1b. More details in determination of $T_{MIT}$ are shown in Fig. S5. As shown in Fig. 1c, the X-ray diffraction patterns of the $Nd_{0.55}Eu_{0.45}NiO_3$ demonstrate their oriented perovskite crystal structure, which transformed to the $Nd_{0.55}Eu_{0.45}NiO_2$ infinite layers after the topotactic reduction, (see more XRD results in Fig. S6). The reduction process may result in reduced crystallinity and interfacial coherency indicated by the broadened XRD peak, as also observed previously in vacuum deposited IL-nickelates[9,10]. The resultant variation in electronic structure, e.g., from $Ni^{3+}$ to $Ni^{1+}$, is further confirmed by synchrotron-based X-ray absorption spectroscopy (XAS) analysis, as results shown in Fig. 7. Figures. 1d-f show the archetypal cross-section morphology for $Nd_{1-x}Eu_xNiO_2/NdGaO_3$ (110) as probed by high-angle annular dark-field (HAADF), where a larger thickness of ~20 nm is observed compared with the vacuum deposited ones (e.g., 6-8 nm)[9,10,22,34]. More representative microscopic morphologies and elementary distributions are demonstrated in Figs. S8-S11. It is also worth noticing that the $Nd_{1-x}Eu_xNiO_3$ precursor coherently grown on the $NdGaO_3$ substrate (see Fig. S12), while the topotactic reduction slightly reduces the crystallinity and coherency of the interface.

**Tunning the Eu-related competition between reentrant and high-$T_c$ superconductivity**

Robust superconducting behaviors are observed for $Nd_{1-x}Eu_xNiO_2$ with $x$=0.25-0.55, with the highest $T_{c,onset}$, $T_{c,50\%}$ and $T_{c,zero}$ emerged at optimum Eu constituents (x=0.35-0.45) reaching ~37.6 K, ~27.3 K and ~22.2 K, respectively. The temperature dependence in resistivity ($\rho$-$T$) of archetypal $Nd_{1-x}Eu_xNiO_2$ are shown in Fig. 2a for optimum doping (x=0.45), underdoping (x=0.3), and overdoping (x=0.52), respectively; while the $\rho$-$T$ tendencies for more Eu-doping constituents are provided in Fig. S13. The optimum doped $Nd_{1-x}Eu_xNiO_2$ (x=0.35-0.45) manifest expected high-$T_c$ superconductivity that is monotonically suppressed by magnetic fields up to 35 T with no signature of reentrance (see Fig. S14). Figs. S14J and S14K further demonstrate its high magnitude of $J_c$ reaching 266 kA/cm$^2$ at 2K, exceeding previous vacuum deposited Nd-Sr and Nd-Eu systems at the same temperature (e.g., $J_c$=170 kA/cm$^2$ for $Nd_{0.8}Sr_{0.2}NiO_2$[9], $J_c$=237 kA/cm$^2$ for $Nd_{0.65}Eu_{0.35}NiO_2$[8]). Only limited reduction in the magnitudes of $J_c$ is observed, even imparting cross-plane magnetic fields up to 9 T. In contrast, the magnetic field reentrant superconducting behaviors reaching zero-resistance are observed for $Nd_{1-x}Eu_xNiO_2$ at both underdoped (x=0.2-0.3) and overdoped (x=0.52-0.6) regions. This is demonstrated by their $\rho$-$B$ tendency in the insets of Fig. 2a (also Fig. S14), and also the crossing in their $\rho$-$T$ tendencies as measured under different magnetic fields in Fig. S15.

The field-reentrant superconductivity emerging at both the underdoped and overdoped boundaries also holds for $Pr_{1-x}Eu_xNiO_2/NdGaO_3$ (110) and $Nd_{1-x}Eu_xNiO_2/(LaAlO_3)_{0.3}(Sr_2TaAlO_6)_{0.7}$ (100), as demonstrated in Fig. S16. This indicates the competition between reentrant and high-$T_c$ superconductivity within Eu-doped IL-nickelates, giving rise to a more complicated superconducting phase diagram ($T_c$-x) compared to the Sr/Ca-doped ones, as shown in Fig. 2c. We highlight that the superconducting reentry emerges exclusively at the superconducting boundaries, but is absent when the high-$T_c$ superconductivity is robust at optimum doping. This unveils the previously underestimated tunability in the quantum criticality by *RE*-4*f* (particularly Eu-4$f^7$)[6], via the likelihood of a magnetic fluctuation modulated pairing in the neighborhoods of quantum phase transitions[11]. For example, in the underdoped region, the superconducting state has energy that is close to many competing states, such as charge order[35], spin order[36] and/or oxygen order[37]. In the overdoped region near a hidden quantum critical point (QCP), the system exhibits a transition from non-Fermi liquid to Fermi liquid behavior[30], triggering a sudden change in electronic state. Hence, potential *RE*-4*f* modulated magnetic fluctuations and the suppression of superconductivity by the competing quantum states[34] may lead to the criticality between reentrant and high-$T_c$ superconductivity[11,36].

Of particular note is the broader superconducting dome extending to larger Eu compositions in $Pr_{1-x}Eu_xNiO_2$ than in $Nd_{1-x}Eu_xNiO_2$, despite the similar magnitudes in their starting point (e.g., x=0.2) and maximum $T_c$ both of ~37.6 K. Further X-ray photoemission spectroscopy analysis as shown in Figs. S18a-c indicate a higher $Eu^{2+}/Eu^{3+}$ ratio for $Nd_{1-x}Eu_xNiO_2$ (e.g., ~1/1) compared to $Pr_{1-x}Eu_xNiO_2$ (e.g., ~2/3). Hence, estimating their practical hole doping range contributed by $Eu^{2+}$ indicates similarities to the previous reports for $Nd_{1-x}Sr_xNiO_2$[30] and $Pr_{1-x}Sr_xNiO_2$[38], as more clearly demonstrated in Figs. S17d and S17e. It is also worth noticing that in other contemporaneous report[6], the reentrant superconducting behavior of $Sm_{0.95-x}Ca_{0.05}Eu_xNiO_2$ via vacuum growth emerges exclusively in the overdoped region, but not the underdoped one. Identifying the underlying reason requires further investigations, as there are fundamental distinctions in both the material composition and methodology of growth between the present work and ref[6]. One of the possibilities lies in the likelihood of the variation in superconducting boundaries via introducing Ca and/or Sm with smaller ionic radius compared to this work. Another possibility that cannot be completely excluded at this moment is potential extrinsic impacts from the microscopic defects or interfacial charge discontinuity, in particular when growing infinite-layers on perovskites substrate. Although a decreasing tendency in cross-plane lattice parameter with x is indicated by the leftwards shift in diffraction peaks of both $Nd_{1-x}Eu_xNiO_2$ and $Pr_{1-x}Eu_xNiO_2$ (see Fig. S6b-d), completely ruling out the above extrinsic impacts yet requires further targeted experimentation.

**Further modulations in the quantum criticality and $T_c$ via multiple *RE*-compositions**

To further elucidate the role of *RE* in competition between high-$T_c$ and reentrant superconductivity, we start with the matrix of optimum doped $Nd_{1-x}Eu_xNiO_2$ (x: 0.35-0.45) without reentrancy and partially substitute Nd by *RE*' (e.g., Pr, Sm, Gd and Dy). The *RE'* substitution elicited corresponding changes in the $T_{MIT}$ of $(Nd_{1-y}RE'_y)_{0.65}Eu_{0.35}NiO_3$ (see Fig. S18) and also lattice constants of $(Nd_{1-y}RE'_y)_{0.65}Eu_{0.35}NiO_2$ (see Fig. S19). Fig. 3a (upper) shows the *ρ-T* tendencies for $(Nd_{0.9}Pr_{0.1})_{0.65}Eu_{0.35}NiO_2$, $(Nd_{0.9}Sm_{0.1})_{0.65}Eu_{0.35}NiO_2$ and $(Nd_{0.9}Dy_{0.1})_{0.65}Eu_{0.35}NiO_2$, where robust superconducting behaviors are observed, e.g., with $T_{c,zero}$ of ~20 K, 18 K and 14 K, respectively. As their *ρ-B* tendencies further demonstrated in the insets and also Figs. S20a-c, these samples display no field-reentrant behaviors. In contrast, the reentrant superconductivity emerges for $(Nd_{1-y}Gd_y)_{0.65}Eu_{0.35}NiO_2$ (y=0.05, 0.1 and 0.2) as demonstrated in Fig. 3a (lower) and also the inset. Their $\rho$-$B$ tendencies measured at more temperatures are further shown in Figs. S20d and S20e. Also, the Gd substitution results in lower $T_{c,zero}$ compared to the ones substituted by other *RE'* (see Fig. S21).

To verify the zero-resistance for the Gd reinforced reentrant superconductivity and its secondary quench, we measured their $\rho$-$B$ tendencies of $(Nd_{0.95}Gd_{0.05})_{0.65}Eu_{0.35}NiO_2$ up to higher magnetic fields at different temperatures, as shown in Fig. 3b. The reentrant superconductivity reaching zero resistances is observed from 5-10 T, 8-13 T and 10-15 T at 1.6 K, 2 K, and 3 K, respectively, afterwards quenching at 30 T, 25 T and 20 T. Further angular dependent magneto-transport measurements via altering the angle ($\theta$) between $H_{ext}$ and *c*-axis indicates strong uniaxial *cross-plane* anisotropy in the superconducting reentrant behavior of $(Nd_{0.95}Gd_{0.05})_{0.65}Eu_{0.35}NiO_2$. Their resistivity is mapped as a function of both magnetic field strength and $\theta$ in Fig. 3c at 2 K, 3 K and 5 K, where more separated regions associated with the low field and reentrant field superconductivity are observed when imparting $H_{ext}$ along *c*-axis ($\theta$=0°). Also, elevating the temperature (e.g., from 2 K to 5 K) results in markedly contraction in the reentrant region. This should be attributed to the higher $H_{c2}$ along the *ab*-plane compared to the *c*-axis, stemming from the Eu-4*f* related anisotropy in orbital depairing[21], as further discussed in Fig. S22. Similar phenomenon was also observed for as-grown $(Nd_{0.9}Gd_{0.1})_{0.65}Eu_{0.35}NiO_2$, $Nd_{0.48}Eu_{0.52}NiO_2$ and $Nd_{0.46}Eu_{0.54}NiO_2$, as demonstrated in Fig. S23. Hence, the Gd reinforced reentrant superconductivity is in consistency to ones observed for Nd-Eu and Sm-Eu systems grown by both vacuum depositions[6-8].

The preferential superconducting reentrancy via introducing Gd over other *RE'* (e.g., Pr, Sm and Dy) within $(Nd_{1-y}RE'_y)_{1-x}Eu_xNiO_2$ was also verified in other compositions (e.g., x=0.35, y=0.2; x=0.45, y=0.2; x=0.5, y=0.1), as demonstrated in Fig. S24. Instead of other proposed mechanisms (e.g., spin-triplet pairing[11] and metamagnetic criticality[39]), the Jaccarino-Peter (J-P) effect[40,41] was more likely to explain the magnetic field induced reentrant superconductivity for Eu-doped IL-nickelates[6-8]. In sight of its half-filled $4f^7$ among $RE^{3+}$ (see details in Table S1)[42,43], introducing $Gd^{3+}$ is expected to enhance the internal exchange field ($H_{ex}$) to partially against the external magnetic field ($H_{ext}$), as illustrated in Fig. 3c. In contrast, the $Eu^{2+}$ (also with $4f^7$) not only generates $H_{ex}$, but also served as hole dopant that simultaneously contributes to transports, giving rise to more complicated impact beyond simply J-P effect.

From a counterpoint compared to Gd, introducing *RE'* (e.g., Pr, Sm and $Pr_{1/2}Sm_{1/2}$) to reduce the magnetism of rare-earth site within $(Nd_{1-y}RE'_y)_{0.65}Eu_{0.35}NiO_2$ sheds a light on further elevating $T_c$ beyond the superconducting dome of the matrix $Nd_{1-x}Eu_xNiO_2$ with $T_c$ (e.g., peaking at ~37.6 K). This was confirmed by higher magnitudes of $T_c$ reaching 40.1 K, 38.1 K and 38.4 K as observed for $(Nd_{0.8}Pr_{0.2})_{0.65}Eu_{0.35}NiO_2$, $(Nd_{0.6}Sm_{0.4})_{0.65}Eu_{0.35}NiO_2$ and $(Nd_{0.4}Pr_{0.3}Sm_{0.3})_{0.65}Eu_{0.35}NiO_2$, respectively, as demonstrated by Figs. S24 and S25. Considering that $Nd^{3+}$ exhibits an ionic radius between $Pr^{3+}$ and $Sm^{3+}$ and a larger magnetic moment than both, the further elevation in $T_c$ is unlikely to be related to rare-earth size effect, but more associated with a weakened lattice magnetism via disturbing *RE*-4*f* spin ordering from *RE*-mixing.

**Elevated $T_c$ coincide with promoted magnetic fluctuation at superconducting state**

Grounded in present large *RE*-compositional diversity, we summarized the $T_c$ plotted as a function of the *c*-axis lattice constants (*c*) for the present Eu-doped IL-nickelates showing robust superconductivity without reentrancy together with the reported ones[10,27] in Fig. 4a. This is further compared to the $T_c$-*c* tendency developed from Sr-/Ca-doped IL-nickelates approaching to optimum doping[30,31,38,44-48], as indicated by the dash line. The further elevation in $T_c$ is clearly observed for the optimum Eu-doped IL-nickelates transcending the Sr-/Ca-doped ones even at similar layer spacing. This unveils an extra pairing strengthening mechanism related to $Eu^{2+}$-4$f^7$, as is further confirmed by performing magneto-electrical transport measurements shown in Fig. S26 to estimate the paring strength summarized in Table S2. It evokes closed associations with potential more complicated spin and orbital interactions with $Eu^{2+}$-4$f^7$, which improves paring via potentially promoted magnetic fluctuation[49], struggling for primacy with the J-P effects.

It is also intriguing to note the abruptly promoted spin ordering below $T_c$ even for the optimum-doped $Nd_{0.6}Eu_{0.4}NiO_2$ with completely shielded reentrant superconductivity, as probed by LE-$\mu$SR, as demonstrated in Fig. 4b. More detailed results are further shown in Figs. S27 and S28. Descending temperature across 20-40 K (consistent to superconducting transition) results in an abrupt increase in the relaxation rate ($\lambda_{ZF}$), accompanied by reductions in the stretch parameter $\beta$ down to a plateau magnitude of ~0.5-0.6 similar to spin-glass[48]. The magnitude of $\lambda_{ZF}$ (e.g., 2.5 $\mu s^{-1}$) in the superconducting state is much larger than that previously observed for the Sr-doped IL-nickelates (e.g., 0.22-0.48 $\mu s^{-1}$)[48]. This illuminates a potential path to further explore associations between the promoted spin ordering in superconducting state and the significant elevation in $T_c$ observed for Eu-doped IL-nickelates (e.g., ~37.6 K) compared to the Sr-doped ones (e.g., ~10 K[9]).

## CONCLUSION

In conclusion, grounded in the high-$p_{O2}$ chemical avenue to effectively grow $(RE_{1-y}RE'_y)_{1-x}Eu_xNiO_2$, we elucidate the enrichment in their superconducting phase diagram by field-reentrant superconducting states emerging at both quantum boundaries, competing for primacy. The quantum criticality between reentrant and high-$T_c$ superconductivity not only hinges on Eu-composition, but also displays associations with the *RE* (*RE*') magnetism, which modulate the exchange field together with $Eu^{2+}$. Furthermore, combining magneto-transport and LE-$\mu$SR measurements unveils the concomitant occurrence of strengthened Cooper pairing and promoted magnetic fluctuation via optimum $Eu^{2+}$-4$f^7$ interactions. This counts for their further elevated $T_c$ beyond the Sr-/Ca-doped systems even at similar infinite layer-spacing, giving rise to higher $T_c$ up to ~40 K and more robust $J_c$ reaching 266 kA/cm$^2$ at 2 K. These discoveries highlight a new freedom from the overlooked *RE*-4*f* orbital and spin interplays that further complicate the superconducting phase diagram of IL-nickelates in terms of quantum phase criticality modulation and pairing strengthening, beyond analogous cuprates with similar *d*-orbital frameworks. Furthermore, the non-vacuum growth of IL-nickelate with high effectiveness holds practical implications, particularly in potential to leverage coated-conductor technologies similar to cuprates.

## METHODS

### Sample preparation

Thin films of $(Nd_{1-y}RE'_y)_{1-x}Eu_xNiO_2$ (*RE'*=Pr, Sm, Gd and Dy) were grown via a high-oxygen-pressure-assisted chemical route, followed by a soft topotactic reduction. Metal nitrate precursors, such as $Nd(NO_3)_3$ and $Eu(NO_3)_3$, as well as the $Ni(CH_3COO)_2$ (≥99.9%, Aladdin) were dissolved in ethylene glycol monomethyl ether (EGME) according to the target stoichiometry. For partial rare-earth substitutions, $Pr(NO_3)_3$, $Sm(NO_3)_3$, $Gd(NO_3)_3$ and $Dy(NO_3)_3$ were co-dissolved correspondingly. The precursor solution was spin-coated onto single-crystal $NdGaO_3$(110) and LSAT(100) substrates and dried in air. As-deposited gel films were crystallized into perovskite-type nickelate precursors by annealing at 700-900 °C under oxygen pressures of 1-20 MPa in a custom-built high-$p_{O2}$ tubular reactor. After cooling to room temperature, the perovskite films were sealed in a vacuum quartz tube together with $CaH_2$ powder and annealed at 260-320 °C for 0.5-4 h for topotactic reduction into the infinite-layer phase. This process simultaneously reduced $Eu^{3+}$ to $Eu^{2+}$, serving as the effective hole dopant. The resulting infinite-layer films had a typical thickness of ~20 nm.

### X-ray diffraction

Crystal structures of the perovskite precursors and reduced infinite-layer nickelate films were characterized using X-ray diffraction (XRD, Bruker D8 Discover) with Cu $K_{\alpha 1}$ radiation ($\lambda$=1.5406 Å). The $\theta$-2$\theta$ scans were collected to identify phase purity, and reciprocal space mapping (RSM) of the (103) reflection was employed to verify the epitaxial quality and in-plane lattice coherence.

### Scanning transmission electron microscopy

The cross-sectional morphologies and interfacial structures of the samples were characterized by scanning transmission electron microscopy (STEM) using an FEI Titan Themis Z microscope operated at 300 kV, equipped with a double aberration corrector and a high-angle annular dark-field (HAADF) detector. The convergence semi-angle for imaging is 25 mrad, and the collection semi-angle is 50-200 mrad for the HAADF imaging. The cross-sectional TEM lamellas were thinned to electron beam transparency at 30 kV by using Carl Zeiss Crossbeam 550L FIB-SEM, followed by the removal of the surface amorphous layer at 2 kV. The obtained HAADF-STEM images revealed atomically sharp interfaces between the $Nd_{1-x}Eu_xNiO_2$ films and $NdGaO_3$ substrates and confirmed the topotactic structural transformation from the perovskite precursor to the infinite-layer phase after reduction. The electron energy-loss spectroscopy (EELS) and energy-dispersive X-ray spectroscopy (EDS) were performed to obtain chemical and electronic-structure information. The EELS spectrum-imaging was acquired with the post-column imaging filter fitted with a K3 direct electron detector. The STEM beam current was set to ~80 pA for spectrum imaging, the pixel dwell time was 1 ms, and the recorded pixel size was 0.5 Å. The EELS energy dispersion was of 0.35 eV per channel. The EDS elemental maps were collected with an acquisition current of ~100 pA. All EELS and EDS datasets were energy- and spatially-calibrated and processed using standard routines; elemental maps shown in the manuscript were optimized in Velox software (Thermo Fisher) with a light Gaussian blur applied for presentation.

**XPS and XAS measurements**

X-ray photoelectron spectroscopy (XPS, Thermo Scientific ESCALAB 250Xi) was used to determine the chemical states of Eu, Nd, and Ni ions, with Al $K_\alpha$ (1486.6 eV) radiation. The binding energy was calibrated using the C 1*s* peak at 284.8 eV. The $Eu^{3+}$:$Eu^{2+}$ ratio was estimated by deconvolution of the Eu $3d_{3/2}$ and $3d_{5/2}$ peaks. The synchrotron-based X-ray absorption spectroscopy (XAS) at the Ni *L*-edge was carried out at the BL08U1A beamline of the Shanghai Synchrotron Radiation Facility (SSRF). Total electron yield (TEY) mode was used to probe the near-surface region, confirming the reduction of Ni valence from +3 in $Nd_{0.65}Eu_{0.35}NiO_3$ to +1 in $Nd_{0.65}Eu_{0.35}NiO_2$.

**Electrical transport measurements**

Electrical transport properties were measured using a Quantum Design Physical Property Measurement System (PPMS). Standard four-probe geometry was used with aluminium ultrasonic wire bonds; contact resistances were typically below a few hundred ohms. Temperature-dependent resistivity ($\rho$-$T$) measurements were performed from 300 K to 2 K under magnetic fields up to 9 T. Angle-dependent transport was measured using a high-precision rotator (angular resolution 0.01°) to determine the anisotropy of reentrant superconductivity. The critical current density ($J_c$) was measured using the Electrical Transport Option (ETO) module of the same PPMS from current-voltage ($I$-$V$) curves in a four-probe configuration.

**High-field magneto-transport**

High-magnetic-field resistance measurements were performed on the WM1 system at the Steady High Magnetic Field Facility, Chinese Academy of Sciences (Hefei). Magnetic fields up to 35 T and temperatures down to 1.6 K were achieved in a custom-built $^4$He cryostat (Physike Instruments). A Keithley 6221 current source and SR830 lock-in amplifiers were used for low-noise data acquisition.

**The $\mu$SR measurements**

The $\mu$SR measurements were performed at the low-energy Muon Facility (LEM) at the $\mu$E4 beam line of the Swiss muon source (S$\mu$S) at Paul Scherrer Institut (PSI) in Villigen, Switzerland. Implantation energies were selected from TRIM.SP simulations to maximise muon stopping rates in the $Nd_{0.6}Eu_{0.4}NiO_2$ films. To exclude the possibility of stray magnetic field during the ZF-$\mu$SR measurements, all the magnets were preliminarily degaussed, and an active field-compensating facility was used. For the wTF-$\mu$SR measurements, the applied magnetic field (i.e., 10 mT) was perpendicular to the muon-spin direction, and the films were cooled in an applied magnetic field down to the base temperature (≈1.6 K). Six pieces of $Nd_{0.6}Eu_{0.4}NiO_2$/LSAT(100) films (each has a dimension of 5×10 $mm^2$) were mounted on the non-magnetic Ag sample holder. The muon spin relaxation of Ag is less than 0.04 $\mu s^{-1}$ at temperatures above 2 K, which is negligible compared with $Nd_{0.6}Eu_{0.4}NiO_2$ films.

## DATA AVAILABILITY

The data that support the findings of this study are available from the corresponding author upon reasonable request.

## DATA

data and video are available at *NSR* online.

## ACKNOWLEDGEMENTS

We thank the BL08U1A beamline and the User Experiment Assist System of the Shanghai Synchrotron Radiation Facility (SSRF) for the assistance in characterizations. We also thank the staff members of the XMCD beamline at the NSRL in Hefei, for providing the technical support and assistance in XAS data collection and analysis. We also thank the WM1 of the Steady High Magnetic Field Facility, Chinese Academy of Sciences, for the assistance on the experiment. The muon measurements were performed at the $\mu$E4 beamline of the Swiss Muon Source S$\mu$S, Paul Scherrer Institute, Villigen, Switzerland. We also acknowledge the support from Dr. Andreas Suter and Dr. Thomas Proksha in PSI Center for Neutron and Muon Sciences, Paul Scherrer Institute, Switzerland. Also, we acknowledge the constructive discussions with Dr. Ivan Božović.

## FUNDING

This work was supported the National Key Research and Development Program of China (2021YFA0718900), the National Natural Science Foundation of China (62474017, 92577108 and 12474001), the Natural Science Foundation of Guangdong Province of China (2025A1515011071) and the Guangdong Provincial Quantum Science Strategic Initiative (GDZX2501006).

## AUTHOR CONTRIBUTIONS

J.C. and J.N. conceived the project. H.H. performed the soft chemical reduction experiments, transport measurements, and XRD characterizations, assisted by Y.B. and T.M.. Y.Z. synthesized the perovskite precursor films assisted by N.C. Y.B. and W.Y. assisted in the transport measurements. H.H., C.X. and Z.W. performed the high-field magneto-transport measurements. S.Y. and B.G. conducted the STEM experiments. H.D performed the XAS experiments assisted by K.M.; T.S. and Z. S. performed the LE-$\mu$SR measurements, while T.S. analysed the data. J.C, J.N. B.G, N.C and K.M. provided constructive experimental supports and discussions. All authors analysed and discussed the data. J.C. wrote the manuscript, assisted by H.H. and J.N, also with input from all authors.

***Conflict of interest statement.*** None declared.

**Figures and captions**

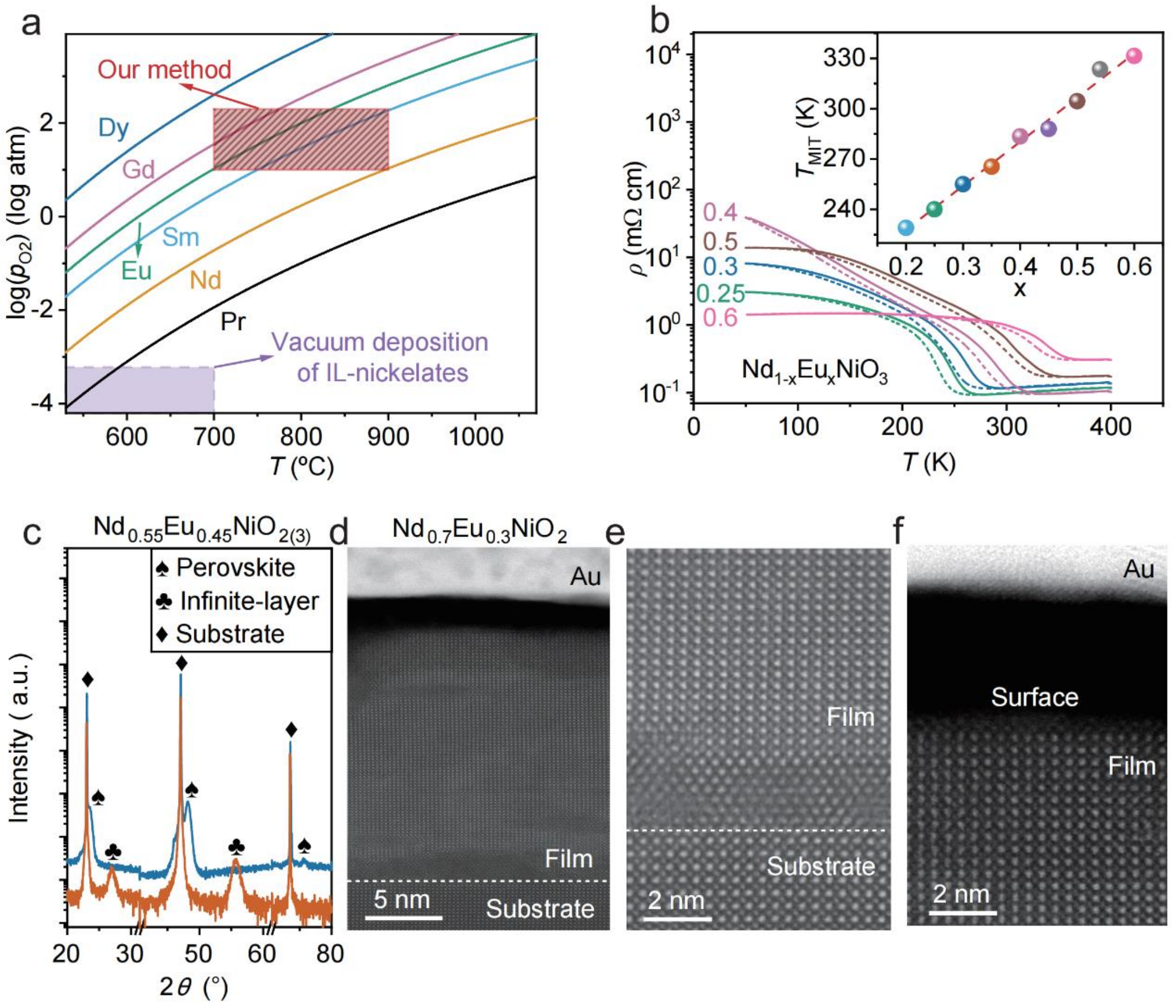

**Fig. 1. High oxygen pressure assisted chemical avenue for infinite-layer nickelates.** (a) The pressure-temperature (*p*-*T*) phase diagram for the formation of perovskite nickelates with different rare-earth elements under high oxygen pressure is depicted. The red region outlines the range of synthesis conditions encompassed by our method, with the slash area indicating the specific parameters employed in this work. The purple region represents the typical synthesis conditions for vacuum deposition of perovskite nickelate thin films. (b) Temperature dependence of resistivity measured for $Nd_{1-x}Eu_xNiO_3$ thin films with various Eu-substituting compositions (*x*), convincing their abrupt metal-insulator transition behavior. The inset shows their metal-insulator transition temperature ($T_{MIT}$) versus *x*. (c) The X-ray diffraction patterns for $Nd_{0.55}Eu_{45}NiO_3$ perovskite precursor and the $Nd_{0.55}Eu_{0.45}NiO_2$ infinite layer. (d) Cross-sectional HAADF-STEM image of $Nd_{0.7}Eu_{0.3}NiO_2$ on $NdGaO_3$(110). (e and f) Magnified HAADF-STEM images of (e) the interfacial region and (f) the film surface.

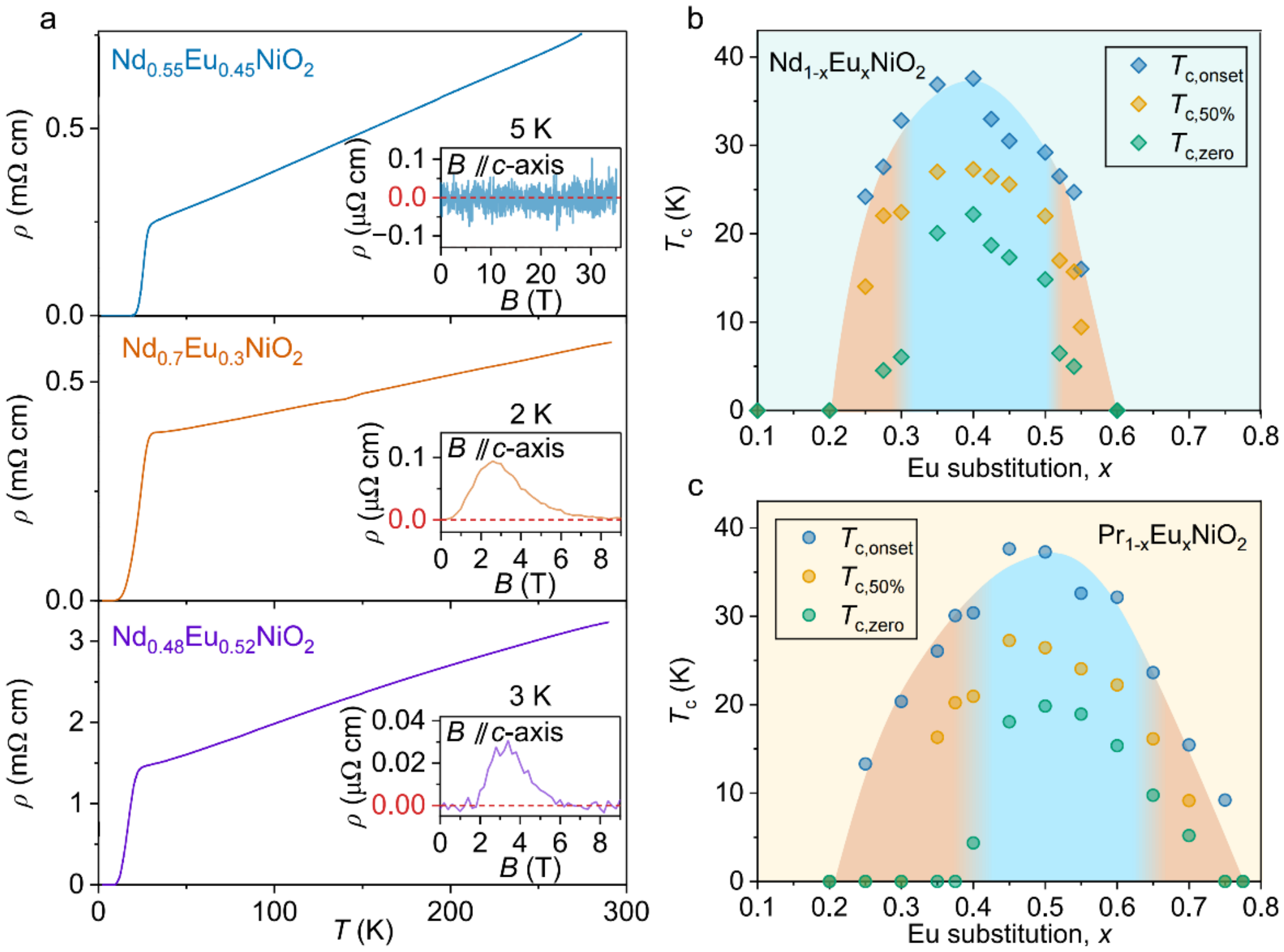

**Fig. 2. Discovery of field reentrant superconductivity at the quantum boundaries of superconducting dome.** (a) The temperature-dependent resistivity ($\rho$-$T$) measured for: the optimally doped $Nd_{0.55}Eu_{0.45}NiO_2$, with the inset showing the corresponding magnetic field dependent resistivity ($\rho$-$B$) at 5 K that demonstrated conventional superconductivity; the underdoped $Nd_{0.7}Eu_{0.3}NiO_2$, with the inset showing $\rho$-$B$ at 2 K that demonstrates reentrance; the overdoped $Nd_{0.48}Eu_{0.52}NiO_2$, with the inset showing $\rho$-$B$ curve at 3 K that demonstrates reentrance. (b and c) Superconducting phase diagram for (b) $Nd_{1-x}Eu_xNiO_2$ and (c) $Pr_{1-x}Eu_xNiO_2$ plotted as a function of Eu substituting composition, $x$. The blue region represents the high-$T_c$ superconducting region without reentry, while the orange-shaded regions represent the emergence of magnetic field-reentrant superconductivity. The onset critical temperature ($T_{c,onset}$) is corresponding to the temperature where the electrical resistivity begins to deviate notably from its normal-state behavior. The midpoint transition temperature ($T_{c,50\%}$) represents the temperature where the resistivity drops to 50% of its normal-state value under zero magnetic field (typically referenced to the value at $T_{c,onset}$). The zero-resistance temperature ($T_{c,zero}$) is defined as the temperature where the electrical resistivity drops below the experimental detection limit. The dome is established from samples at more Eu-substituting compositions, as shown by Figs. S13 and S16.

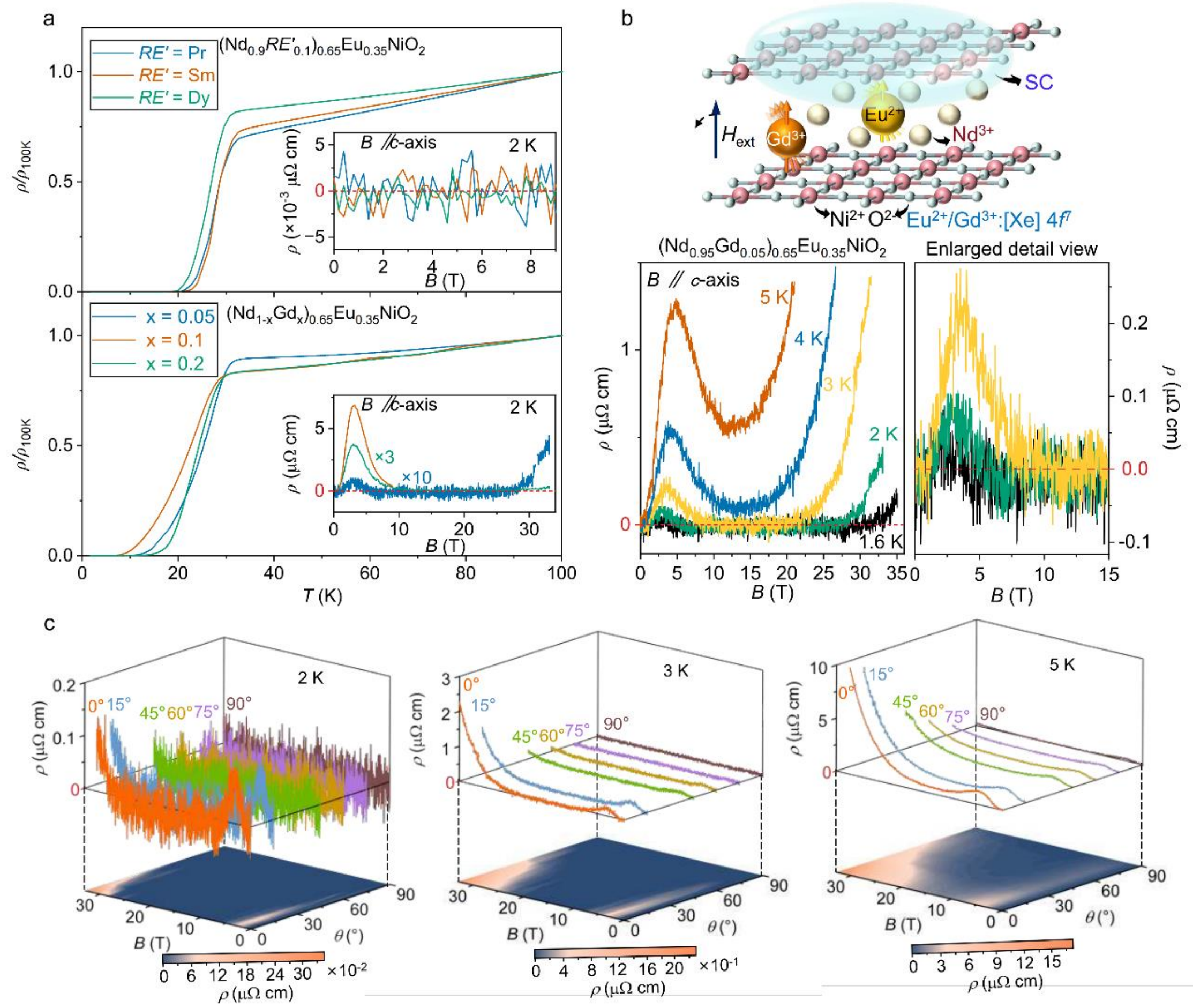

**Fig. 3. Further modification in criticality with field-reentrant superconductivity via introducing other rare-earth substituents.** (a) The normalized temperature dependent resistivity ($\rho$-$T$) measured for $(Nd_{0.9}RE'_{0.1})_{0.65}Eu_{0.35}NiO_2$ ($RE'$=Pr, Sm and Dy), displaying robust conventional superconductivity. The inset shows their corresponding magnetic field dependent resistivity ($\rho$-$B$) measured at 2 K. The normalized $\rho$-$T$ as measured for $(Nd_{1-y}Gd_y)_{0.65}Eu_{0.35}NiO_2$ (y=0.05, 0.1 and 0.2), showing clear field-reentrant superconductivity. The inset shows their corresponding $\rho$-$B$ measured at 2 K. (b) Schematic illustration of the proposed mechanism: the $4f^7$ moments of $Gd^{3+}$ located at deeper energy level generate an internal exchange field ($H_{ex}$) that partially compensates the external magnetic field ($H_{ext}$) via the Jaccarino-Peter effect[40], giving rise to reentrant superconductivity. $\rho$-$B$ measurements of $(Nd_{0.95}Gd_{0.05})_{0.65}Eu_{0.35}NiO_2$ up to higher magnetic fields at different temperature. (c) Resistivity of $(Nd_{0.95}Gd_{0.05})_{0.65}Eu_{0.35}NiO_2$ measured for various angles between the external magnetic field and $c$-axis at: 2 K, 3 K and 5 K. The bottom projection shows the colored mapping of the resistivity, where the deep blue regions of zero resistance (reentrant superconductivity) emerging at high fields when $H_{ext}$ is near the $c$-axis ($\theta \approx 0°$), revealing strong uniaxial anisotropy. In addition, elevating the temperature progressively suppresses the reentrant superconducting phase, constraining it within a narrow range of angles and fields.

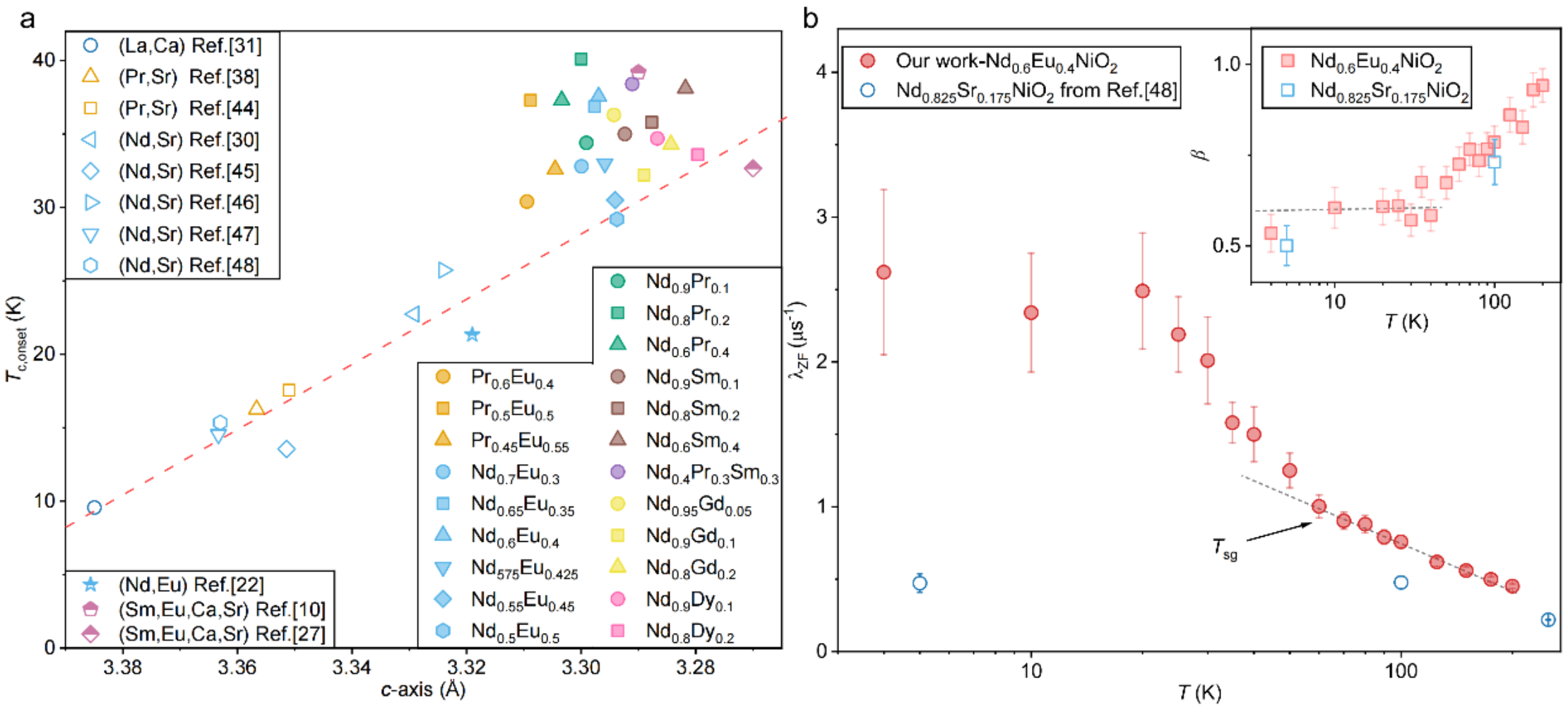

**Fig. 4. Further elevation in $T_c$ concomitant with promoted magnetic fluctuations in Eu-doped infinite-layer nickelates.** (a)The $T_{c,onset}$ plotted as a function of the $c$-axis lattice constant for the present and reported[10,22,27] Eu-doped infinite-layer nickelates without reentrancy together with ones doped by Sr/Ca[30,31,38,44-48]. The dashed line is a guide to the $T_c$-$c$ trend fitted from the Sr-/Ca-doped systems. (b) The zero-field muon spin relaxation rate ($\lambda_{ZF}$) fitted from the zero-field (ZF-) $\mu$SR spectra plotted as a function of temperature for the optimum-doped $Nd_{0.6}Eu_{0.4}NiO_2$ films grown on the LSAT (100) substrates, compared to the reported ones for $Nd_{1-x}Sr_xNiO_2$ (hollow dots)[48]. Descending temperature across $T_c$ results in abrupt elevation in $\lambda_{ZF}$, indicating promoted spin ordering and magnetic fluctuation. The inset demonstrates the temperature dependence of the stretch exponent $\beta$ (solid dots), showing similar temperature dependence compared with the previous report for $Nd_{1-x}Sr_xNiO_2$ (hollow dots)[48]. The magnitude of $\beta$ saturates to ~0.5-0.6 below 50 K, indicating a possible spin glass-like state, as $T_{sg}$ marked by the arrow.